# QUADRANTIDS FILAMENTS MODELING

Rosaev A.E.

OAO NPC NEDRA, Jaroslavl, Russia

Numeric integration of orbits of particles along mean orbit of Quadrantid meteor stream is done at time span 20000 years. Orbits are subdivided on several classes by their evolution type. A very complex dynamical behavior is detected. About 20% of modeled particles escape stream: this fact point on that stream cannot be long-live and have a source within 5000 years.

After that, Quadrantid filaments dynamics are studied. By comparison of different authors data, 7 independent filaments are selected. Only four of them can have age more than 500 year. Others are unstable or have a small age. Only few Quadrantid filaments can be continuous and observable at all longitudes. Other filaments consist of separate clumps, which occurs number of meteor phenomenon in separate years and non active at different epoch.

A complex view of filament dynamical evolution argues in favor multistage ejection particles into a stream at different epoch.

**1. Introduction**

The Quadrantids, one of the more active of the annual meteor showers, is unusual for its strong but brief maximum within a broader background of activity. The Quadrantids move in a short-period orbit with a period of revolution of about 5 years, at a high inclination of $70^o$ and approach the orbit of Jupiter. The range of orbital parameters is very broad. Orbits of 57 radio-observed Quadrantid, studied in [1], have the same nodal longitude and perihelion distance, but very different semiaxis. The values of orbital elements variation with time are estimated in [1].

Until recently, no parent with a similar orbit had been observed and previous investigators concluded that the stream was quite old, with the stream's recent appearance and sharp peak attributed to a fortuitous convergence of meteoroid orbits. The discovery of the near-Earth Asteroid 2003 EH1 on an orbit very similar to that of the Quadrantids provides, probably, the parent body of this stream [2].

As noted in [3], **multi-stage origin for the Quadrantids is most like**. The central core of the stream today probably originates with 2003 EH1 as first proposed by Jenniskens and Marsden [2]. Jenniskens [4] suggest that the **most likely time period for the formation of the** central portion of the stream approximately **500 years ago**. The first observation of the stream having been likely made in 1835. Based on this fact, authors [3] suggest the basic model having 2003 EH1 as the parent of the core of the stream is correct, but that the formation age is closer to **1750–1800 AD**. However, the interval of possible observation Quadrantides, as a meteor stream in Earth's sky, is 1870-2190 yrs.[5] A number of Quadrantids numeric integration shows that at 1100-1800 years before present a stream has a lower inclination, much smaller perihelion distance. Possible, it means, that Quadrantid meteor **stream was formed 1690-1300 year before present**. **[5].** Most suitable time for main part of meteor ejection is when inclination and perihelion distance of orbit are small (eccentricity is maximal). It is true as well as for cometary or collisional mechanism for ejection. However, some part of stream may be younger.

In **[6]** by applying a computerized stream search procedure (utilizing the D-criterion) to photographic meteor orbits, 60 Quadrantid meteors were found in the IAU MDC catalogue. Applying a strict limit for the D criterion, D=0.03, five distinct filaments within the stream were found which may correspond to different ejection events of the stream formation. Filaments 1 and 2 undergo the same orbital evolution as 2003 EH1 and so they are most likely associated with the asteroid [6], nature of other filaments still unknown. Early, five filaments in stream (with slightly different elements) are described in **[7]**, where 118 orbits from meteor catalogue are in use.

## 2. Mean orbit dynamical evolution classification

Quadrantids mean elements, slightly different for different authors (see **[6]**). Our choice is an averaged value, most close to Lindblad data, table 1, where elements of most probable parent body (**2003 EH1**) is given. Actually observed Quadrantids orbits are in a wide range of phase space, between 2:1 and 3:1 resonances with Jupiter.

Table 1. Quadrantids mean elements and possible parent body

|  | *a* | *e* | *i* | ω | Ω |
|---|---|---|---|---|---|
| **Quadrantids** | 3.05 | 0.677 | 71.4 | 171.0 | 283.0 |
| **2003 EH1** | 3.128 | 0.619 | 70.8 | 171.4 | 282.9 |

First, we integrate 180 clones, placed equidistantly in true anomaly along orbit, during 20000 years. The perturbations from 7 planets (except Mercury) on fixed circular (elliptic) orbit. The Runge-Kutta 7-8 order algorithm is used. Step of integration is selected equal 0.1 day for all integrations.

There are six main classes of orbits for Quadrantids mean orbits from table 1. As it may be expected, there are few areas of concentration and disappearance of particles along an orbit. The presence of escaped orbit can lead to varying a stream activity from year to year.

I. Most numerous class (66 clones or 37%) – chaotic bounded orbits, when semimajor axis display chaotic variations, but in certain range. The eccentricity evolutions in nearly always cases hold a periodic character.

II. Escape/Collisions class. A very unstable orbits with hyperbolic (or collisions with Sun) final orbit. About 15% of modeled particles escape stream on time interval 20000 year. The lifetime of particles in according parts of orbit is not large 10-15 thousands year.

III. Class of orbits with tendency of decrease of semimajor axis (30 clones or more 16%). In two cases it leads to escape particles from stream. The most probable Quadrantid parent body (2003 EH1) is included in this class.

IV. Similar class III but with increasing semimajor axis (7%), most of them escape stream (IV e class).

V. A very interesting class with (temporary) trapping in regime with periodic variations of semimajor axis. There are about 19% modeled orbits.

VI. Class of orbit with destroyed periodicity in *a* (5%, escape resonance).

## 3. Quadrantid filaments dynamic

It is known a few strands or filaments of stream, for different authors by Wu and Williams **[7]** and by Porubcan and Kornos **[6],** their show non-coincidence too (see table 2). It is interesting to compare dynamical evolution of filaments from [6] and [7] to their real distinguishes. There are some addition questions: which filaments are more stable, how an activity of filaments can change in time and can some filaments have a common origin.

To study these problems, we made series of numeric integrations of filaments orbits. First, we integrate 18 clones, placed equidistantly in true anomaly along orbit of each filament, during 2700 years. Than we integrate 36 clones for each filament with small variation of initial positions and velocities to study dispersion of orbits under planetary perturbations. Wiegert and Brown [3] study dynamical evolution of 2003 EH1 clones. As it followed from their results, significant divergence of clone's orbits starts since about 500 years ago.

Table 2. Quadrantids filaments

|       | a     | e     | i    | ω     | Ω     | Reference |
|-------|-------|-------|------|-------|-------|-----------|
| FIP   | 2.867 | 0.657 | 72.0 | 176.8 | 282.7 | [6]       |
| FIIP  | 2.931 | 0.666 | 72.2 | 170.3 | 283.2 | [6]       |
| FIIIP | 2.735 | 0.641 | 70.8 | 173.3 | 283.3 | [6]       |
| FIVP  | 3.176 | 0.693 | 70.9 | 168.0 | 283.3 | [6]       |
| FVP   | 3.131 | 0.686 | 72.9 | 174.5 | 283.0 | [6]       |
| FIW   | 2.98  | 0.673 | 71.7 | 170.7 | 282.1 | [7]       |
| FIIW  | 2.83  | 0.654 | 70.8 | 167.7 | 282.1 | [7]       |
| FIIIW | 2.35  | 0.585 | 71.4 | 164.3 | 281.7 | [7]       |
| FIVW  | 4.10  | 0.761 | 72.0 | 173.7 | 282.5 | [7]       |
| FVW   | 3.33  | 0.708 | 71.5 | 168.6 | 282.3 | [7]       |

Results of our research are in following (table 3).

F1P. Range semiaxis variation 2.75-2.915. Eccentricity (max e=0.96) and inclinations evolution curves have small width and extremum (min i=15 degree) about 1700 years ago (**fig.1**). Possible stable filament at all longitudes and has common origin. **Dispersion semimajor axis with time is relative slow (fig.2): possible age of this filament up to 1700 year.**

F2P. Range semiaxis variation 2.85-3.015. Eccentricity (max e=0.96) and inclinations evolution curves have broad width and extremum about 1700 years ago. Range minimum in inclination (15 degree) is wide: 1400-2100 years ago (**fig.3**). **Fast spreading close orbits (fig.4) show, that this filament cannot be longlive, not more than 500 year old**. Maybe, this filament consists of a few clumps. Related to III class in our classifications.

F3P. Range semiaxis variation 2.727-2.777. Eccentricity (max e=0.965) and inclinations (17 degree) evolution curves are very narrow and extremum about 2200 years ago (**fig.1**). **Possible stable filament at all longitudes, and all particles have common origin. Dispersion semimajor axis with time is very slow (fig.2): possible age of this filament up to 2200 year.**

F4P. Range semiaxis variation 2.95-3.4. Eccentricity (max e=0.97) and inclinations evolution curves have broad width and extremum about 1700 years ago. Range minimum in inclination (10 degree) is wide: 1200-2000 years ago (**fig.1**). **Significant dispersion semimajor axis begins since 1500 years ago (fig.2). Related to V class in our classifications.**

In general, results of our modeling for filaments F1P, F3P and F4P are in agreement with [6].

F5P. Range semiaxis variation up to hyperbolic orbits. Eccentricity (max e>1) and inclinations evolution curves have broad width no extremum. Range minimum in inclination (10 degree) is wide and have few minima: 1500-2000 years ago. **For this filament we obtain more chaotic evolution than in [6]. Maybe, this filament consists of a few clumps. In our classification, filament has II class.**

F1W. Range semiaxis variation 2.92-3.2. Eccentricity (max e=0.96) and inclinations evolution curves have broad width and extremum about 1700 years ago. Two minimums in inclination (15 degree) are wide: 1200-1500 and about 2000 years ago. **In according with [7], this filament is strongly perturbed by Jupiter. Approximately coincide with filament F4P.**

F2W. Range semiaxis variation 2.80-2.875. Eccentricity (max e=0.96) and inclinations evolution curves have broad width and extremum about 2200 years ago. Ranges of minimums in inclination (20 degree) are wide: 1700-1900 and about 2200-2600 years ago. **Ranges for real meteors minimal inclinations in [7] is 2200-2600 year. It is strongly advice that this filament is a clump, and not extended over all longitudes. Approximately coincide with filament F2P.**

F3W. Range semiaxis variation 2.354-2.376. Eccentricity and inclinations evolution curves have small width and extremum more than 2700 years ago. **Possible stable filament at all longitudes and has common origin. In general, results of our modeling are in agreement with [7].**

**F4W.** Range semiaxis variation 4.09-4.175. Eccentricity and inclinations evolution curves have no extremum. Significant pericenter rotation is detected. **Possible, it is not Quadrantids as it noted in [7].**

**F5W.** Range semiaxis variation 3.1-3.52. **Our modeling confirms a very chaotic (but bounded) evolution for this filament as is noted in [7]. In our classification, filament has I class.**

Finally, we can conclude, that filament I by Wu and Williams (F1W) and filament IV by Porubcan and Kornos (F4P) with great possibility are the same filament. Note, that it is ton followed from mean elements (table 2). With smaller possibility, F2P and F2W are the same too.

Only filaments F1P, F3P and F3W are different, spread along all longitudes and longlive.

Table 3.

| N | Filament | Range $a$ | Epoch of minimal $i$ | Notes | Maximal possible age, yr | eccentricity variation period, yr | Dynamic. class |
|---|---|---|---|---|---|---|---|
| 1 | F1P | 2.75-2.915 | 1700 | Real filament | 1700 | 3200 | IV |
| 2 | F2P (F2W) | 2.85-3.015 2.80-2.875 | 1400-2100 | Clumps or young | 500 | 3500 | III |
| 3 | F3P | 2.727-2.777 | 2200 | Real filament | 2000 | 5000 | I |
| 4 | F4P (F1W) | 2.95-3.4 2.92-3.2 | 1200-2000 | Clumps or young | 1500 (250) | 4200 | V |
| 5 | F5P | - | 1500-2000 | unstable | <500 | - | II |
| 6 | F3W | 2.354-2.376 | >2700 | Real filament | 2700 | 10000 | I |
| 7 | F5W | 3.1-3.52 | 1700-2700 | clumps | - | - | I |

### 3. Conclusions

Numeric integration of orbits of particles along mean orbit of Quadrantid meteor stream is done at time span 20000 years. Orbits are subdivided on several classes by their evolution type. A very complex dynamical behavior is detected. About 20% of modeled particles escape stream: this fact point on that stream cannot be long-live and have a source within 5000 years.

After that, Quadrantid filaments dynamics are studied. By comparison of different authors data, 7 independent filaments are selected. Only four of them can have age more than 500 year. Others are unstable or have a small age. Only few Quadrantid filaments can be continuous and observable at all longitudes. Other filaments consist of separate clumps, which occurs number of meteor phenomenon in separate years and non active at different epoch. These clumps have different dynamical history. A complex view of filament dynamical evolution argues in favor multistage ejection particles into a stream at different epoch.

### Acknowledgement

Author sincerely thanks professor I.P. Williams for a very important discussion and notes.

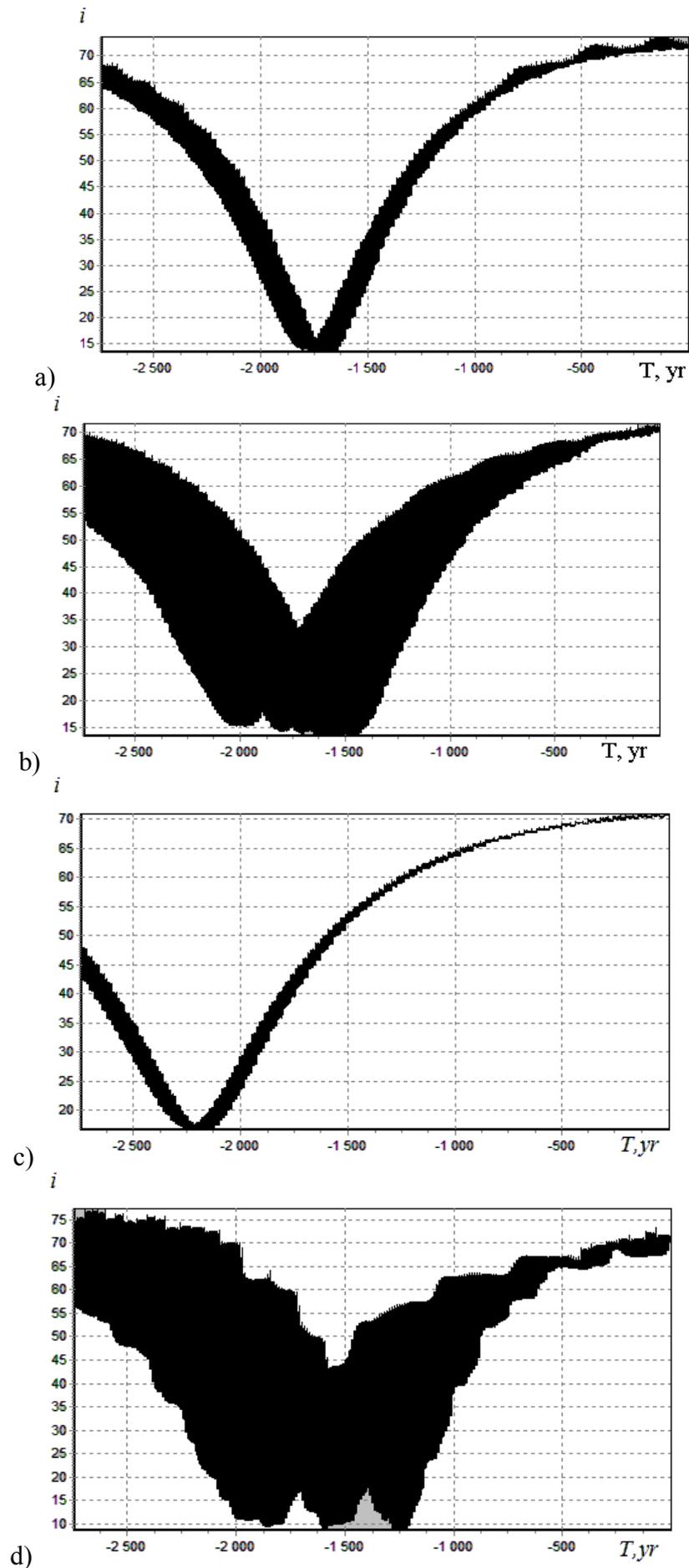

Fig.1 Filaments clones inclination evolution (a – F1P, b – F2P, c – F3P, d –F4P)

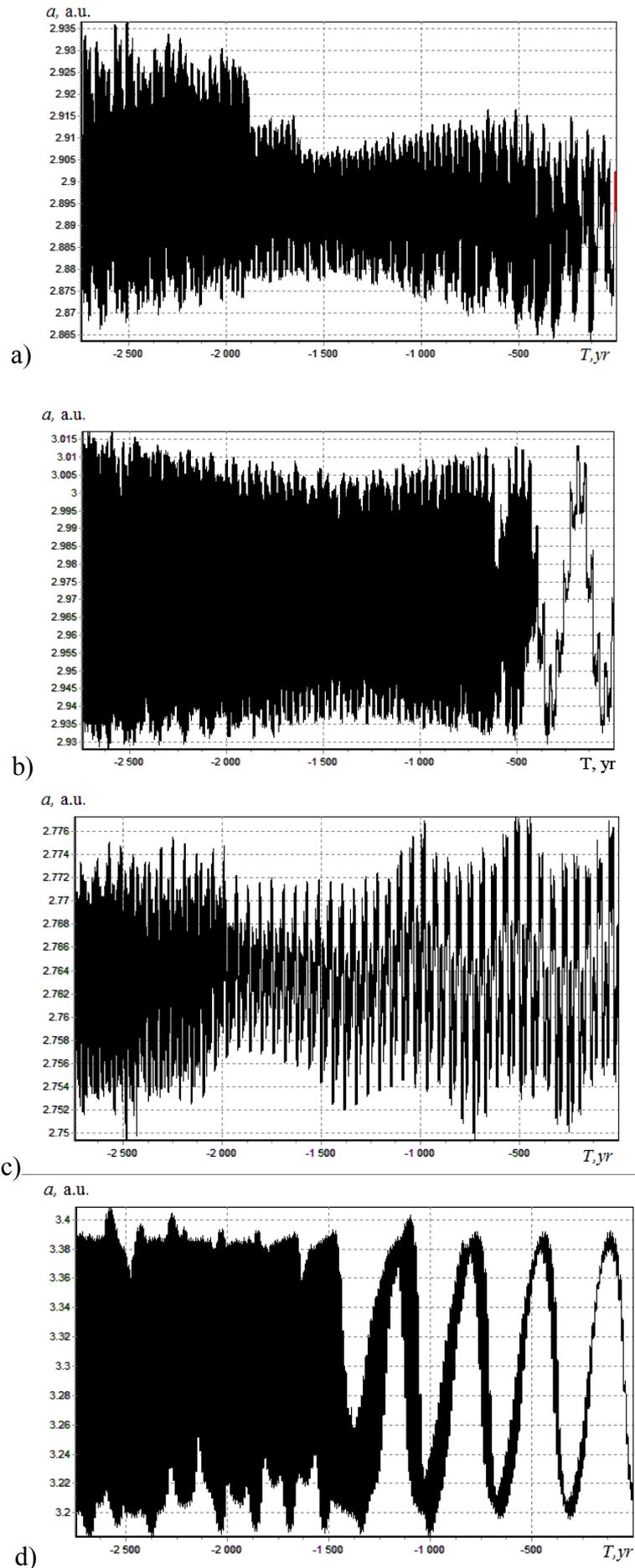

Fig.2. Filament clones semimajor axis evolution (a – F1P, b – F2P, c – F3P, d –F4P)